\documentclass{article}

\usepackage{arxiv}

\usepackage[utf8]{inputenc} 
\usepackage[T1]{fontenc}    
\usepackage{hyperref}       
\usepackage{url}            
\usepackage{booktabs}       
\usepackage{amsfonts}       
\usepackage{nicefrac}       
\usepackage{microtype}      
\usepackage{lipsum}
\usepackage{amsthm}
\usepackage{graphicx}
 
\theoremstyle{definition}
\newtheorem{definition}{Definition}
 
\theoremstyle{remark}

\usepackage[compact]{titlesec}
\titlespacing{\section}{24pt}{*0}{*0}
\titlespacing{\subsection}{0pt}{*0}{*0}
\titlespacing{\subsubsection}{0pt}{*0}{*0}
 
\title{Online Disinformation \\ and the Role of Wikipedia}

\author{Diego Saez-Trumper \\ Wikimedia Foundation \\ diego@wikimedia.org}

\begin{document}
\maketitle

\begin{abstract}
The aim of this study is to find key areas of research that can be useful to fight against disinformation on Wikipedia. To address this problem we perform a literature review trying to answer three main questions: \emph{(i)} What is disinformation? \emph{(ii)}What are the most popular mechanisms to spread online disinformation?  and \emph{(iii)} Which are the mechanisms that are currently being used to fight against disinformation?. 

In all these three questions we take first a general approach, considering studies from different areas such as journalism and communications, sociology, philosophy, information and political sciences. And comparing those studies with the current situation on the Wikipedia ecosystem.

We found that disinformation can be defined as non-accidentally misleading information that is likely to create false beliefs. While the exact definition of misinformation varies across different authors, they tend to agree that disinformation is different from other types of misinformation, because it requires the intention of deceiving the receiver.  A more actionable way to scope disinformation is to define it as a problem of information quality. In Wikipedia quality of information is mainly controlled by the policies of neutral point of view and verifiability.

The mechanisms used to spread online disinformation include the coordinated action of online brigades, the usage of bots, and other techniques to create fake content. Underresouced topics and communities are especially vulnerable to such attacks. The usage of sock-puppets  is one of the most important  problems for Wikipedia. 

The techniques used to fight against information on the internet, include manual fact checking done by agencies and communities, as well as automatic techniques to assess the quality and credibility of a given information. Machine learning approaches can be fully automatic or can be used as tools by human fact checkers. Wikipedia and especially Wikidata play double role here, because they are used by automatic methods as ground-truth to determine the credibility of an information, and at the same time (and for that reason) they are  the target of many attacks. Currently, the main defense of Wikimedia projects against fake news is the work done by  community members and especially by \emph{patrollers}, that use mixed techniques to detect and control disinformation campaigns on Wikipedia.

We conclude that in order to keep Wikipedia as free as possible from disinformation, it’s necessary to help \emph{patrollers} to early detect disinformation and assess the credibility of external sources. More research is needed to develop tools that use state-of-the-art machine learning techniques to detect potentially dangerous content, empowering \emph{patrollers} to deal with attacks that are becoming more complex and sophisticated. 

\end{abstract}

\section{Introduction}

\begin{figure}[h]
\caption{Number of page views for the \emph{Fake News} article in the English Wikipedia. The interest about this topic has grown significantly (note that the y-axis is logarithmic) since the end of 2016. }
\centering
\includegraphics[width=0.98\textwidth]{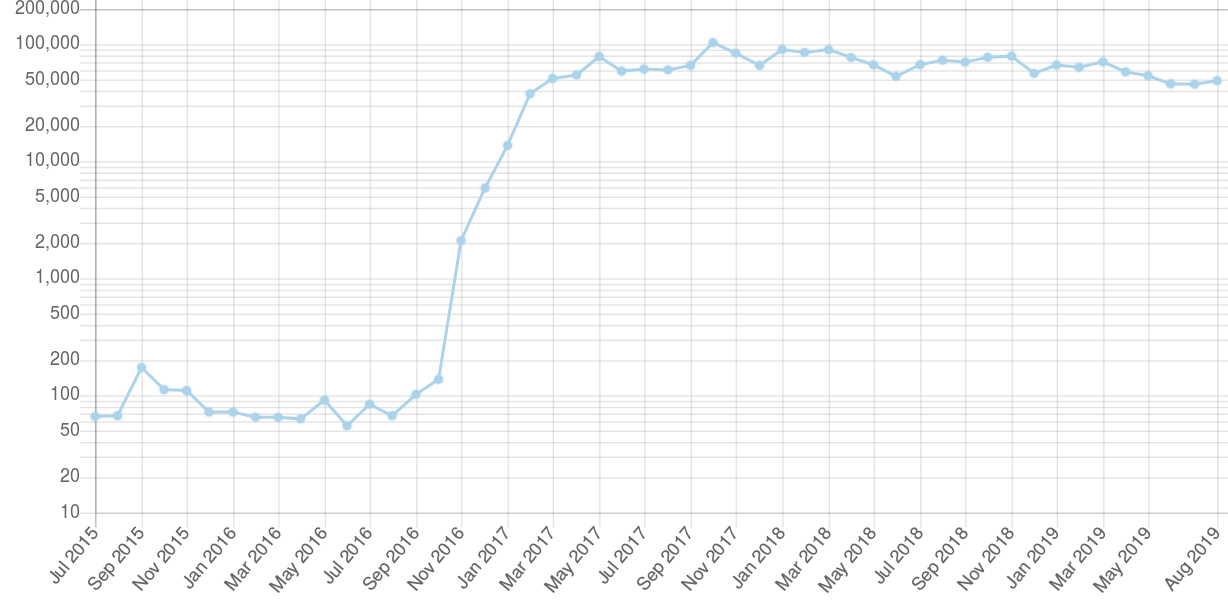}
\label{fig:pageviews}
\end{figure}

Concepts like disinformation, fake news and post-truth has become popular in the last years (see Figure \ref{fig:pageviews}). Even thought that usage and spread of false information with political purposes is not new, the changes in technology can have an important impact in the way that people receive and process information~\cite{lessons1942}. The rise of user generated content (UGC) brings a lot of opportunities for getting more diverse, faster and complete information, but also creates new challenges in terms of quality and reliability of information. But have have changed in the last few years?  While during the first years of growth of UGC platforms most of the problems where related to quality of information (spread of hoaxes and rumors)~\cite{castillo2011information,messner2011legitimizing}, more recently social media has become a battle-field, having  potential impact on political decisions and elections~\cite{dimitrova2018social}. While researchers and  western media have paid special attention to the impact of online propaganda during the  2016 US presidential elections~\cite{guess2018selective,grinberg2019fake}, or the Brexit referendum in the United Kingdom~\cite{bastos2019brexit,cadwalladr2017great}, other similar cases have been registered in countries such as Brazil~\cite{brazilWhatsApp}, India~\cite{McLaughlin2018Dec} and Kenya~\cite{Kenyansf43:online}.  What all the aforementioned cases have in common is the existence of groups of people organized to disseminate false information for electoral purposes. Techniques used to introduce such disinformation might include the usage of bots, online activists and \emph{click farms}~\cite{marwick2017media,wasapBrazilClicks}. 
Moreover, the creation of filter bubbles~\cite{pariser2011filter} where people are only exposed to ideas and sources that reinforce their previous beliefs, the concept  of fake-news has been used as a rhetorical device to discredit information that affects a given point of view~\cite{ribeiro2017everything}\footnote{This is idea is well described by the title of the cited paper: "Everything I Disagree With is\# FakeNews":}.

Given the relevance and complexity of this problem, online disinformation has been studied from a wide spectrum of disciplines such as philosophy~\cite{floridi2013philosophy,fetzer2004disinformation,fallis2009conceptual}, communication~\cite{marwick2017media}, political~\cite{tucker2018social} and computer science~\cite{cardoso2019can,zhou2018fake}. In this work we give an overview of these studies, trying to answer these three questions:
\begin{enumerate}
    \item What is disinformation?
    \item What are the most popular mechanism to spread online disinformation ?
    \item  Which are the mechanisms that are currently being used to fight against disinformation?
\end{enumerate}

Given the extension, diversity and complexity of the literature in this field, we focus our efforts in getting actionable answers for these three questions, and putting special emphasis on the role of  Wikimedia projects in the information/disinformation ecosystem.  Differently from other platforms and online communities, Wikipedia (and Wikidata) have a double role in this space, being at the same time part of the battle-field were different visions tries to reflect their ideas, but also being the ground-truth~\cite{thorne2018fever} (or reference) used by automatic and manual fact checking systems.

The remainder of this document is organized as follows: first we mention other surveys and literature reviews in the area of disinformation, describing the main differences with our work, next we dedicate a section for each of the three questions listed above and  finally we discuss potential lines of research that will be useful to keep Wikipedia as free as possible from disinformation and depict our conclusions. 

\section{Related Work}

Because this paper is a literature review, we consider as related work other studies having similar purposes. As mentioned earlier, the phenomena of disinformation has been studied from different areas of knowledge. In the field of philosophy we found a comprehensive essay by Fallis~\cite{fallis2015disinformation}, discussing the definition of disinformation. We build on top of that essay in section 3. In that section we contrast Fallis outputs with an interesting work by Wardle \& Derakhshan~\cite{wardle2017information}, that describes disinformation from a communication and political science point of view. From the computer science perspective, a very detailed and complete survey has been done by Zhou and Zafrani in 2018~\cite{zhou2018fake}, we build on top of their work adding some of the last work in the field and adding the Wikipedia perspective. 

All the studies mentioned in this section are complementary to our work, but they are focused in a specific area of knowledge, and don’t put emphasis on the importance of the Wikimedia projects in the disinformation ecosystem. They are good material in case that the reader wants to go deeper in any of the aforementioned fields.  

\section{What is disinformation?}

According to Fallis~\cite{fallis2015disinformation} disinformation has these three characteristics: 
\begin{itemize}
    \item Disinformation is a type of information.
    \item Disinformation is a type of misleading information; that is  information that is likely to create false beliefs.
    \item Disinformation is non accidentally misleading information.
\end{itemize}

Summarizing, the three claims above, we can define:
\theoremstyle{definition}
\begin{definition}{Disinformation}
 is non-accidentally misleading information that is likely to create false beliefs.
\end{definition}

Similarly, Wardle and Derakhshan says: “Disinformation is when false information is knowingly shared to cause harm"~\cite{wardle2017information}, adding definitions two other relevant concepts:

\theoremstyle{definition}
\begin{definition}{Misinformation}
 is when false information is shared, but no harm is meant.
\end{definition}

Interestingly, they~\cite{wardle2017information} also introduce the concept of \textit{mal-information}, where the focus is not in faking the information, but in publishing sensitive private information with the aim causing harm.
\theoremstyle{definition}
\begin{definition}{Mal-information }
is when genuine information is shared to cause harm, often by moving information designed to stay private into the public sphere.
\end{definition}

An analogous definition for \emph{fake news} is given Zhou and Zafrani: “Fake news is intentionally and verifiably false news published by a news outlet”~\cite{zhou2018fake}, for them disinformation is a related concept to fake news, that implies inauthentic content and bad intention. Interestingly, they introduce a set of other related concepts such as: Maliciously false news (false and bad intention), Satiric News (without bad intention), Misinformation (where authenticity is unknown but there is no bad intention) and rumors (where the authenticity and intentions are unknown), and categorize them according to the information authenticity and emisors' intention (see Table \ref{tab:intention}). 
\begin{table}[!t]
\begin{center}
\caption{Following the taxonomy proposed by  Zhou and Zafrani~\cite{zhou2018fake} and adding the definitions proposed by Wardle and Derakhshan \cite{wardle2017information}, Fallis \cite{fallis2015disinformation}, a summary of the different concepts related with disinformation, according two their Authenticity (ie. veracity) and intention.}
\begin{tabular}{|l|l|l|}
\hline
\textbf{}                & \textbf{Authenticity} & \textbf{Intention} \\ \hline
\textbf{Disinformation}  & False                 & Bad                \\ \hline
\textbf{Misinformation}  & False                 & Unknown            \\ \hline
\textbf{Mal-Information} & True                  & Bad                \\ \hline
\textbf{Fake News}       & False                 & Bad                \\ \hline
\textbf{Satire News}     & False                 & Not Bad            \\ \hline
\textbf{Imposter Content}           & False               & Unknown            \\ \hline
\textbf{Fabricated Content}           & False               & Bad            \\ \hline
\textbf{Manipulated Content}           & Unknown               & Bad            \\ \hline
\textbf{Rumor}           & Unknown               & Unknown            \\ \hline
\end{tabular}
\label{tab:intention}
\end{center}
\end{table}

Wardle and Derakhshan~\cite{wardle2017information}, also describe 7 types of mis/disinformation:

\begin{itemize}

  \item \emph{Satire Content}:  Without intention to cause harm.
   \item \emph{Misleading Content}: misleading use of  information  to frame an issue.
  \item \emph{Imposter Content}: impersonating genuines sources.
  \item \emph{Fabricated Content}: completely fake content designed to deceive.
  \item \emph{False Connection}: When headlines, visuals or captions don’t support the content.
  \item \emph{False Context}: when genuine content is shared with false contextual information.
  \item \emph{Manipulated Content}: when genuine or fake content is manipulated to deceive.
\end{itemize}

While these definitions are consistent among them, they are not directly actionable because they assume the existence of false beliefs (and as a consequence the existence of true beliefs) that are verifiable,  and the need of asses the intention behind a given piece of information. In many cases the authenticity of a given information is verifiable, but this is not the case for controversial topics~\cite{borra2014contropedia}. 

A more actionable approach is proposed by Castillo~\cite{castillo2016big}, where  mis/disformation, and the assessment of veracity, is scoped as problem of quality of information, proposing a set of practices and metrics to evaluate the quality of a given piece of information.

\subsection{Wikipedia: Verifiability and Neutral Point of View}

Considering the information quality approach to define understand disinformation, in the context of Wikipedia we can consider the 3 core content policies as the axis to assess the quality of a given information. These policies are:

\begin{itemize}

    \item Neutral Point of View (NPOV): ``which means representing fairly, proportionately, and, as far as possible, without editorial bias, all of the significant views that have been published by reliable sources on a topic."~\cite{Wikipedi81:neutral} 

    \item Verifiability: ``means other people using the encyclopedia can check that the information comes from a reliable source."~\cite{Wikipedi87:veri}

    \item No original research: ``Wikipedia articles must not contain original research”.~\cite{Wikipedi36:online}

\end{itemize}

While the first-one address the aforementioned problem of having a unique truth, the second one addresses the problem of assessing the authenticity of given content.  

These 3 policies are actionable, allowing to deal with many of the practical problems of writing encyclopedic content in a collective manner~\cite{arazy2006wisdom}. However, from an epistemic perspective, they transfer the responsibility of assessing the quality and credibility of a content to third parties named reliable sources.  In practice, this helps the Wikimedia movement to write Wikipedia, but it can’t avoid the existence of controversies~\cite{borra2014contropedia} and edit wars~\cite{sumi2011edit,Wikipedi57:online}. Even though some studies shows that the existence of such disputes might end up in producing higher quality articles~\cite{shi2019wisdom}. 

\section{Which mechanism are used to spread online disinformation?}

\begin{table}[]
\centering
\caption{Social attacks classified by the type of weakness they exploit.}
\label{tab:my-table}
\begin{tabular}{|l|p{4cm}|p{5cm}|}
\hline
\textbf{Weakness exploited} & \textbf{Description}                                              & \textbf{Example}                                                                                                                               \\ \hline
Social System               & When reputation systems are hacked to introduce disinformation.   & Use bots or sock-puppets to over-represent an opinion or confirm a false information.                                                          \\ \hline
Lack of Information         & When the lack of information is used to introduce disinformation. & Spread disinformation during on-going events like natural disasters or manipulate search engines results in topics without enough information. \\ \hline
\end{tabular}
\label{tab:socialvstec}
\end{table}

In order to understand the mechanisms used for spread misinformation we split them in two groups: social and technological attacks (for details see Table \ref{tab:socialvstec}). We consider social attacks those mechanisms that exploits weakness in  information systems but without requiring programming or coding knowledge. For example, social attacks will include  mechanisms such as \emph{sock-puppeting}, and \emph{click farms}. Social attacks can be done by individuals or groups of people, but tends to be more effective when kmore people is involved. On the other hand technological attacks, will include all kind of mechanism that requires coding skills, such as the creation of bots, or \emph{deepfake} techniques~\cite{guera2018deepfake}. While these attacks can be done by groups of people, they are usually designed to increase the impact of individual users in a community. In reality, both types of attacks can be combined generating hybrid mechanisms that could be more difficult to detect~\cite{wasapBrazilClicks}

\subsection{Social Attacks}

Within the social attacks we can differentiate two types, the ones that explodes the weakness in social systems, and others that explodes the weakness of the information itself.

\subsubsection{Exploiting the weaknesses of social systems}.

Individuals or groups of people can exploit the weakness of information systems to promote their ideas or to deliberately spread disinformation. Marwick and Lewis have studied the behavior of hate groups to manipulate online information, finding some common patterns in such groups: ‘’they act as organized brigades which engage in a cooperative competition to increase harm to their victims, reinforcing social dominance over marginalized groups"~\cite{marwick2017media}. While these cybermobs are usually related with attacks to specific people or communities~\cite{krumsiek2017cyber}, they can also be used to promote disinformation in a broader sense~\cite{citron2019cyber}. 

Another technique of using crowds to deceive online communities is the usage of click farms~\cite{whatisClick}. While this concept was originally referring to people (low) paid to click on advertisements, it has also been described as ‘’An undercover operation in which individuals fraudulently interact with a website to artificially boost the status of a client’s website, product or service.”~\cite{ClickFarm2} Recently, during the 2018 Brazilian elections, the usage of click farms to spread disinformation has been reported~\cite{wasapBrazilClicks}.

On the other hand, there are individuals that creates multiple online identities to increase their influence in online communities, this usually known as \emph{sock-puppeting}~\cite{sockpuppets} or \emph{sybil attacks}~\cite{sybilAtaccks}. Sock puppets can be different purposes from avoiding bans~\cite{poland2016haters}, submit multiple votes in online pooling systems (Ballot stuffing)~\cite{friedman2007manipulation}, to make point of view to look ridiculous~\cite{orita20074} or to give fake or artificial appearance of support (meatpuppets)~\cite{zhang2013online}. 
\subsubsection{Exploiting the lack of information}.

However, it is not necessary to create multiple accounts or coordinate groups of people to introduce disinformation. For example, during natural disasters single users can introduce rumors (or disinformation, depending on the intention) in platforms such as Twitter that will receive similar amount of attention than real information, even thought that community might refute  those tweets~\cite{mendoza2010twitter}.

Moreover, attackers can exploit the lack of information in a given topic to manipulate the results retrieved by video recommender systems~\cite{ribeiro2019auditing} or the search engines. This kind of weakness of information are called \emph{data voids}~\cite{datavoids}.

\subsection{Technological Attacks}

As defined above  we refer as technological attacks the ones that requires coding skills. The most popular attacks on this field are related with the usage of social bots. In this context we define social bots as software agents that acts in social media trying to influence the course of discussion~\cite{stieglitz2017social}. 
“Bots often imitate or replace a human user's behavior and they typically they do repetitive tasks, and they can do them much faster than human users could”~\cite{Cloudflare}. 
There are several reports on the usage of bots in Twitter to manipulate that social network~\cite{howard2016bots,howard2016botsUK}. 
Just to mention some examples the bots has been used to interfere in elections in Brazil~\cite{arnaudo2017computational}, Japan~\cite{schafer2017japan}, USA~\cite{bessi2016social,howard2018algorithms}, Germany~\cite{brachten2017strategies} and the UK~\cite{howard2016bots}. 
For example in the USA bots were used to influence the social media discussion on the presidential debates. Similar with meatpuppets, bots can be used to create a fake idea of support for an idea or a person, sending messages to support an idea, or creating \textit{human-like} accounts to increase the reputation (ex. number of followers) of a given person, this usually known as Astroturfing~\cite{stieglitz2017social}\footnote{For a detailed categorization of social bots please refer to Stieglitz et al work~\cite{stieglitz2017social}.}. 

But beyond bots, there are new techniques to introduce disinformation.  \emph{Deepfake} is the usage of  generative adversarial networks to create composed videos that looks real~\cite{maras2019determining}. There are examples of fake videos where Barack Obama insults Donald Trump that looks credible~\cite{Schwartz2018Nov}. While there are no massive reports on the usage of this kind of videos on social media yet, they open a big challenge both for human fact checkers and machines to filter such content~\cite{guera2018deepfake,koopman2018detection,li2018exposing}. Also, companies like Facebook are showing concern about this issue~\cite{facebookdeepfakechallege}.

\subsection{Attacks on Wikipedia}
\label{subsec:attacksOnWikipedia}

Within the same Wikipedia community we can find several reports of possible disinformation social attacks. From interest groups trying to impose their views and narratives in some political~\cite{israelWikipedia} or religious topics~\cite{scientology}, until the finding of large groups of paid editors\cite{orangemoody}. Moreover, there are claims of interests groups trying to overtake a full Wikipedia project\footnote{In this context we call Wikipedia Project to a Wikipedia in a given language.}~\cite{azwiki}. Wikimedians have responded this attacks investigating the reported complaints and banning users that don’t comply the movement policies. An interesting case has been reported in the Bulgarian Wikinews community, where they have voted to close Bulgarian Wikinews as they view that there is insufficient community time to combat disinformation  and propaganda there\footnote{At the time of publication of this work (September 2019), the proposal was still under study.}~\cite{bulgaria}. 

Not only the Wikimedians cares about the disinformation on Wikipedia. The press is also looking to this problem. Similarly with the examples above, newspapers have published about political groups overtaking full projects, trying to modify a set of historical articles  to impose their own narrative and views~\cite{Sampson2015DecCroatia}. Moreover, some journalists have been looking to the discussion pages of some controversial articles, questioning the community decisions about which information to include or exclude there~\cite{Brandom2018Mar}. This is interesting because it shows at the same time one weakness of Wikipedia (specialists on a given topic probably having some biases), but also the strength of having an accountable system to understand when and how an (dis)information has been added or removed from Wikipedia. 

Also, the western mass-media is usually reporting disinformation attacks presumably being organized or supported by foreign governments~\cite{Harding2018Apr,Shubber2017Oct}. On the other hand there is no documentation on mass-media about the participation of western governments in disinformation campaigns on Wikipedia, although that the  massive surveillance programs executed by those governments might have affected Wikipedia's readers behavior~\cite{penney2016chilling}.

While journalists and Wikipedians are scanning Wikipedia to prevent disinformation, the relationship between these two actors can also be a weakness. \emph{Circular reporting} or \emph{Citogenesis} in Wikipedia is a problem where some piece of (dis)information is introduced on Wikipedia and later being used as a reference for an external source (with or without citing Wikipedia). If that source is considered reliable, later it can be used as a reference to support the original piece of (dis)information introduced on Wikipedia~\cite{Harrison2019Mar,}. While good practices from  Wikipedia editors (not adding statements without references), and from journalists (using Wikipedia as a third-party source~\cite{citingWikipeda}), there are several cases of circular reporting involving Wikipedia~\cite{circularList}.

\begin{table}[!t]
\centering
\caption{Summary of the most popular mechanism to spread online disinformation.}
\label{tab:attacks}
\begin{tabular}{|l|p{6cm}|l|p{3cm}|}
\hline
\textbf{Mechanism} & \textbf{Description} &\textbf{Type} & \textbf{Wikipedia's Vulnerability}  \\ \hline
Bots               & Software used to automatize the spread of messages, generating the idea that of a lot people is given an specific opinion or interest about a topic & Technical  & Low \\ \hline
Sock-puppets       & Multiple  Online identities used for purposes of deception.                                     & Social & Medium \\ \hline
Web Brigades    & A set of users coordinated to introduce fake content by exploiting the weakness of communities and systems. & Social &  High                           \\ \hline
Click farms    & Where a large group of low-paid workers are hired   to perform some micro-tasks to deceive online systems.     & Social  & Medium                                                          \\ \hline
Deepfake          & AI a technique for human image synthesis that can be used to create fake videos of celebrities or notable people.                       & Technical  & Medium  \\ \hline
Data Voids         & Exploiting missing data to manipulate search results             & Social  & Medium                                                                     \\ \hline
Circular reporting & A situation where a piece of information appears to come from multiple independent sources, but in reality comes from only one source.    & Social & High \\ \hline
\end{tabular}
\end{table}

\subsubsection{Vulnerability of Wikipedia}
Considering the cases described above, social attacks such as the case of Web Brigades and Click farms can be considered two of the most sensitive to  Wikipedia.  The success of such attacks would be related with the volume attackers (paid or volunteers) respect to size of the existing community, thus projects smaller amount of (good-faith) volunteers might be more sensitive to such attacks. More traditional attacks such as sock-puppets are well-known~\cite{sockPupetry} and not  solved problem,  however due to the nature and reach of those attacks, they arguably might have a more localized impact, not being able to overtake a full project. Similarly, the usage of \textit{data voids} for introducing disinformation, will be directly related with the community capacity to  cover all the topics, but the harm created will be again focused on specific topics without enough content. For previous experience (see section \ref{subsec:attacksOnWikipedia}) circular reporting is still a big problem in the Wikipedia context. 

Technological attacks will depend a lot in quality, sophistication and development of such techniques. However, currently with the current status of such technologies, and considering the expertise of Wikipedia patrollers (see Section~\ref{subsubsec:patrollers}) such attacks don’t seem to most dangerous in the present times. Nevertheless, it’s important to keep tracking the evolution of those technologies and generate awareness about them within the Wikipedia community.

\section{How to fight Online Disinformation}

The main way to fight against disinformation is to check the credibility of information, in the field of news this is usually known as fact checking. For the analysis we can split fact checking strategies in two main groups: manual fact checking, and automatic fact checking. The former refers to techniques where humans looks to a given information and evaluate its credibility.  In the latter, machine learning algorithms do this task. In practice, both approaches work together, with human evaluators relaying in software to perform their tasks, or with machines learning from human annotated data. 

Fact checking strategies can also be classified by the action taken by the platform used to spread that information, while in some cases the content can be filtered or deleted, in other cases platforms will add warnings with contextual information about the source or the content itself~\cite{Hatmaker2016Dec,googleFact}. 

\subsection{Manual Fact Checking}
Manual fact checking can be done by communities, like in Wikipedia, or by specialized entities known as fact checking agencies. For example, Facebook hires a large number of fact checking agencies around the world~\cite{facebookhiring} (see Table \ref{tab:factCheckingCats}).

\begin{table}[]
\centering
\caption{Categories used fact checking agencies hired by Facebook to tag content credibility~\cite{facebookCat}.
}
\label{tab:factCheckingCats}
\begin{tabular}{|l|p{7cm}|}
\hline
\textbf{Category} & \textbf{Description}                                                                                                          \\ \hline
False             & The primary claim(s) of the content are factually inaccurate.                                                                 \\ \hline
True              & The primary claim(s) of the content are factually accurate.                                                                   \\ \hline
Mixture           & The claim(s) of the content are a mix of accurate and inaccurate.                                                             \\ \hline
False Headline    & The primary claim(s) of the article body content are true, but the primary claim within the headline is factually inaccurate. \\ \hline
Not eligible      & The content contains a claim that is not verifiable.                                                                          \\ \hline
Satire            & The content is posted by a Page or domain that is a known satire publication.                                                 \\ \hline
Opinion           & The content advocates for ideas and draws conclusions based on the interpretation of facts and data.                          \\ \hline
Prank Generator   & Websites that allow users to create their own “prank” news stories.                                                           \\ \hline
\end{tabular}
\end{table}

There are also websites dedicated to fact checking, like Politifact\footnote{politifact.com} focused in USA politics, or  Snopes\footnote{snopes.com} dedicated to internet rumors in general. 

There are also machine learning approaches that intend to gather information for facilitating the work of human fact checkers. For example, by creating search tools that allows fact checkers to find diverse version of the same fact~\cite{shang2018investigating}; finding suspicious behavior on news spread through Whatsapp~\cite{resende2018system}; or to assess of the political leaning of news source~\cite{ribeiro2018media}.

\subsection{Automatic Fact Checking}

The main problem with manual fact checking is that requires human intervention making difficult to scale to huge amounts of information. That is one of the main reasons to fully automate the process using Natural Language Processing techniques.  Automatic fact checking has two main steps: fact extraction and fact checking. In the first one, given a piece of information, the idea is to extract the facts and/ or claims contained there, this requires to extract the entities associated with such information and a timestamp~\cite{getoor2012entity}; in the second one that claims are contrasted with a trustable knowledge base~\cite{thorne2018fever}. 

Several datasets and challenges has been released during the last years, for example: The Fake news challenge\footnote{http://www.fakenewschallenge.org/}, FEVER\footnote{http://fever.ai} and the clickbait challenge\footnote{https://www.clickbait-challenge.org/}. Those datasets consists of a set of news or piece of text containing claims, and human labels assessing the credibility of such information. Currently, most competitive solutions relies on the usage of deep neural networks~\cite{thorne2018fact}. 

A detailed taxonomy of automatic fact checking techniques can be found in the work proposed by Zhou and Zafrani~\cite{zhou2018fake}.

\subsection{Fact Checking and Wikipedia}

There to components to be analyzed when discussing about fighting against disinformation and Wikipedia. One is how Wikipedia editors check the content added to the online encyclopedia, and another is how third-parties uses Wikipedia to fight disinformation. 

\subsubsection{Fighting against  disinformation within Wikipedia}
\label{subsubsec:patrollers}
As mentioned earlier, the mechanism to fight against disinformation in Wikipedia is approached as an information quality problem, that relies on the usage of reliable sources. In practice, the quality control is community process, where a specialized users called \emph{patrollers}, watches for the compliance of the community norms. These users look for recent changes in Wikipedia articles, check the creation of new pages, or detect vandalism, among other things. Their  work, includes some tools such as anti-vandalism bots and  watchlists (a tool for tracking recent changes on articles)~\cite{jmopatrolling}. There are other tools like ORES that allows to automatically assess the quality of every single article revision~\cite{halfaker2015artificial}. Also NLP approaches have been proposed to detect unsupported statements~\cite{redi2019citation}. However, currently the involvement of machine learning techniques on fighting disinformation is low, and process relies mainly manual procedures~\cite{jmopatrolling}.  
 
\subsubsection{Third-parties using Wikipedia to fight against disinformation}

Wikipedia has been considered a reliable source for fact checking, due its NPOV policy~\cite{nakashole2014language}. Also  Wikidata it is  key resource for doing entity linking and recognition~\cite{spitz2016state}. Algorithms can use Wikidata to extract entities learning its characteristics and the relationship among them, and use them for fact checking tasks~\cite{thorne2018fever}. 

But there are other ways to use Wikipedia for fighting against disinformation. One of the most interesting is to use Wikipedia to provide contextual information to the users~\cite{platformsUsingWikipedia}. For example, Youtube has announced the usage Wikipedia to provide “information cues” on conspiracy theory videos. The idea is simple, when a user is watching a video about a given topic, the system will provide Wikipedia-based information about that topic, assuming this information will be reliable, although that Wikimedia Foundation Executive Director has been  skeptical declaring  that "Frankly, we don't want you to blindly trust us. Sure, we’re mostly accurate - but not always! We want you to read Wikipedia with a critical eye"~\cite{Youtube}. A slightly different approach has been taken by Facebook, they are using Wikipedia to provide information about the published of a given content~\cite{facebookWikipedia}.

\section{Discussion and Final remarks}
We found that Wikipedia plays an important role in the disinformation ecosystems, both as battle-field were attackers try to introduce disinformation but also as a reliable source for machines and humans to contrast disinformation. We  have defined two types of attacks social and technological, finding that Wikipedia usually suffers more from the former than the latter. However, the rise of sophisticated techniques such as deepfake or the growth of click farms, impose new challenges for Wikipedia editors, that currently are not extensively using AI tools to fight disinformation. Also, most of the patrolling work is focusing on detecting misbehavior for single users, but without enough tools to look for coordinated attacks including multiple users. Moreover, the manual fact checking done by Wikipedia editors it is difficult to scale in projects without enough volunteers, increasing the risks of attacks in smaller or under-resourced communities. In those contexts is where machine learning techniques can be extremely helpful. 

During this literature review we have not found studies on disinformation attacks on multiple projects. Currently, we don’t know if attacks occurs similarly in different Wikipedia languages.  Can we train a machine learning algorithms with knowledge generated in the English Wikinews project, to apply it the Bulgarian one? Which will be the implications and biases introduced by such an approach? Moreover, given that smaller languages usually have less NLP resources, can we directly apply methods that will work well in the best resourced languages?

But also we have find research than can be adapted and tested on the Wikipedia context. For example, automatically assigning a credibility score for third-party sources, or assessing their political leaning. 

In general, we can say that the current manual approaches in Wikipedia are working correctly in well resourced communities, but the complexity and amount of attacks might increase in the future, therefore, it is important to help the communities, especially the smaller ones, with state-of-the-art techniques that can simplify their work and amply their results. 

\bibliographystyle{acm}  
\bibliography{references}  

\begin{thebibliography}{10}

\bibitem{arazy2006wisdom}
{\sc Arazy, O., Morgan, W., and Patterson, R.}
\newblock Wisdom of the crowds: Decentralized knowledge construction in
  wikipedia.
\newblock In {\em 16th Annual Workshop on Information Technologies \& Systems
  (WITS) Paper\/} (2006).

\bibitem{arnaudo2017computational}
{\sc Arnaudo, D.}
\newblock Computational propaganda in brazil: Social bots during elections.
\newblock {\em Project on Computational Propaganda 8\/} (2017).

\bibitem{bastos2019brexit}
{\sc Bastos, M.~T., and Mercea, D.}
\newblock The brexit botnet and user-generated hyperpartisan news.
\newblock {\em Social Science Computer Review 37}, 1 (2019), 38--54.

\bibitem{bessi2016social}
{\sc Bessi, A., and Ferrara, E.}
\newblock Social bots distort the 2016 us presidential election online
  discussion.
\newblock {\em First Monday 21}, 11-7 (2016).

\bibitem{borra2014contropedia}
{\sc Borra, E., Weltevrede, E., Ciuccarelli, P., Kaltenbrunner, A., Laniado,
  D., Magni, G., Mauri, M., Rogers, R., Venturini, T., et~al.}
\newblock Contropedia-the analysis and visualization of controversies in
  wikipedia articles.
\newblock In {\em OpenSym\/} (2014), pp.~34--1.

\bibitem{brachten2017strategies}
{\sc Brachten, F., Stieglitz, S., Hofeditz, L., Kloppenborg, K., and Reimann,
  A.}
\newblock Strategies and influence of social bots in a 2017 german state
  election-a case study on twitter.
\newblock {\em arXiv preprint arXiv:1710.07562\/} (2017).

\bibitem{Brandom2018Mar}
{\sc Brandom, R.}
\newblock {How gun buffs took over Wikipedia{'}s AR-15 page}.
\newblock {\em Verge\/} (Mar 2018).

\bibitem{cadwalladr2017great}
{\sc Cadwalladr, C.}
\newblock The great british brexit robbery: how our democracy was hijacked.
\newblock {\em The Guardian 7\/} (2017).

\bibitem{cardoso2019can}
{\sc Cardoso Durier~da Silva, F., Vieira, R., and Garcia, A.~C.}
\newblock Can machines learn to detect fake news? a survey focused on social
  media.
\newblock In {\em Proceedings of the 52nd Hawaii International Conference on
  System Sciences\/} (2019).

\bibitem{ClickFarm2}
{\sc Carr, S.}
\newblock {What Is a Click Farm? Uncovering The Dark Truth Behind Fake Likes}.
\newblock {\em PPC Protect\/} (Sep 2019).

\bibitem{castillo2016big}
{\sc Castillo, C.}
\newblock {\em Big crisis data: social media in disasters and time-critical
  situations}.
\newblock Cambridge University Press, 2016.

\bibitem{castillo2011information}
{\sc Castillo, C., Mendoza, M., and Poblete, B.}
\newblock Information credibility on twitter.
\newblock In {\em Proceedings of the 20th international conference on World
  wide web\/} (2011), ACM, pp.~675--684.

\bibitem{citron2019cyber}
{\sc Citron, D.~K.}
\newblock Cyber mobs, disinformation, and death videos: The internet as it is
  (and as it should be).
\newblock {\em Michigan Law Review, Forthcoming\/} (2019).

\bibitem{Cloudflare}
{\sc Cloudflare}.
\newblock {What Is a Bot? {$\vert$} Bot Definition {$\vert$} Cloudflare}, Sep
  2019.
\newblock [Online; accessed 23. Sep. 2019].

\bibitem{platformsUsingWikipedia}
{\sc Cohen, N.}
\newblock {Perspective {$\vert$} Conspiracy videos? Fake news? Enter Wikipedia,
  the {`}good cop{'} of the Internet}.
\newblock {\em Washington Post\/} (Apr 2018).

\bibitem{dimitrova2018social}
{\sc Dimitrova, D.~V., and Matthes, J.}
\newblock Social media in political campaigning around the world: Theoretical
  and methodological challenges, 2018.

\bibitem{facebookdeepfakechallege}
{\sc {Facebook}}.
\newblock {Creating a data set and a challenge for deepfakes}, Sep 2019.
\newblock [Online; accessed 23. Sep. 2019].

\bibitem{facebookCat}
{\sc Facebook}.
\newblock {Facebook Media and Publisher Help Center}, Sep 2019.
\newblock [Online; accessed 23. Sep. 2019].

\bibitem{fallis2009conceptual}
{\sc Fallis, D.}
\newblock A conceptual analysis of disinformation. iconference.
\newblock {\em Chapel Hill, NC, California, USA\/} (2009).

\bibitem{fallis2015disinformation}
{\sc Fallis, D.}
\newblock What is disinformation?
\newblock {\em library trends 63}, 3 (2015), 401--426.

\bibitem{fetzer2004disinformation}
{\sc Fetzer, J.~H.}
\newblock Disinformation: The use of false information.
\newblock {\em Minds and Machines 14}, 2 (2004), 231--240.

\bibitem{Youtube}
{\sc Field, M.}
\newblock {YouTube will use Wikipedia to tackle fake news - but failed to tell
  Wikipedia}, Sep 2019.
\newblock [Online; accessed 23. Sep. 2019].

\bibitem{floridi2013philosophy}
{\sc Floridi, L.}
\newblock {\em The philosophy of information}.
\newblock OUP Oxford, 2013.

\bibitem{friedman2007manipulation}
{\sc Friedman, E., Resnick, P., and Sami, R.}
\newblock Manipulation-resistant reputation systems.
\newblock {\em Algorithmic Game Theory 677\/} (2007).

\bibitem{getoor2012entity}
{\sc Getoor, L., and Machanavajjhala, A.}
\newblock Entity resolution: theory, practice \& open challenges.
\newblock {\em Proceedings of the VLDB Endowment 5}, 12 (2012), 2018--2019.

\bibitem{datavoids}
{\sc Golebiewski, M., and Boyd, D.}
\newblock {\em Data Voids: Where Missing Data Can Easily Be Exploited}.
\newblock Data \& Society, 2018.

\bibitem{googleFact}
{\sc Google}.
\newblock {See fact checks in search results - Google Search Help}, Sep 2019.
\newblock [Online; accessed 23. Sep. 2019].

\bibitem{grinberg2019fake}
{\sc Grinberg, N., Joseph, K., Friedland, L., Swire-Thompson, B., and Lazer,
  D.}
\newblock Fake news on twitter during the 2016 us presidential election.
\newblock {\em Science 363}, 6425 (2019), 374--378.

\bibitem{guera2018deepfake}
{\sc G{\"u}era, D., and Delp, E.~J.}
\newblock Deepfake video detection using recurrent neural networks.
\newblock In {\em 2018 15th IEEE International Conference on Advanced Video and
  Signal Based Surveillance (AVSS)\/} (2018), IEEE, pp.~1--6.

\bibitem{guess2018selective}
{\sc Guess, A., Nyhan, B., and Reifler, J.}
\newblock Selective exposure to misinformation: Evidence from the consumption
  of fake news during the 2016 us presidential campaign.
\newblock {\em European Research Council 9\/} (2018).

\bibitem{halfaker2015artificial}
{\sc Halfaker, A., and Taraborelli, D.}
\newblock Artificial intelligence service “ores” gives wikipedians x-ray
  specs to see through bad edits, 2015.

\bibitem{Harding2018Apr}
{\sc Harding, L.}
\newblock {Ex-Trump aide Paul Manafort approved 'black ops' to help Ukraine
  president}.
\newblock {\em the Guardian\/} (Apr 2018).

\bibitem{Harrison2019Mar}
{\sc Harrison, S.}
\newblock {The Dizzying Problem of Citationless Wikipedia
  {\textquotedblleft}Facts{\textquotedblright} That Take On a Life of Their
  Own}.
\newblock {\em Slate Magazine\/} (Mar 2019).

\bibitem{Hatmaker2016Dec}
{\sc Hatmaker, T., and Constine, J.}
\newblock {Reports of a Facebook fake news detector are apparently a plugin}.
\newblock {\em TechCrunch\/} (Dec 2016).

\bibitem{howard2016bots}
{\sc Howard, P., Kollanyi, B., and Woolley, S.~C.}
\newblock Bots and automation over twitter during the third us presidential
  debate.

\bibitem{howard2016botsUK}
{\sc Howard, P.~N., and Kollanyi, B.}
\newblock Bots,\# strongerin, and\# brexit: computational propaganda during the
  uk-eu referendum.
\newblock {\em Available at SSRN 2798311\/} (2016).

\bibitem{howard2018algorithms}
{\sc Howard, P.~N., Woolley, S., and Calo, R.}
\newblock Algorithms, bots, and political communication in the us 2016
  election: The challenge of automated political communication for election law
  and administration.
\newblock {\em Journal of information technology \& politics 15}, 2 (2018),
  81--93.

\bibitem{facebookWikipedia}
{\sc Hughes, T., Smith, J., and Leavitt, A.}
\newblock {Helping People Better Assess the Stories They See in News Feed with
  the Context Button {$\vert$} Facebook Newsroom}, Sep 2019.
\newblock [Online; accessed 23. Sep. 2019].

\bibitem{koopman2018detection}
{\sc Koopman, M., Rodriguez, A.~M., and Geradts, Z.}
\newblock Detection of deepfake video manipulation.
\newblock In {\em Conference: IMVIP\/} (2018).

\bibitem{krumsiek2017cyber}
{\sc Krumsiek, A.}
\newblock {\em Cyber Mobs: Destructive Online Communities}.
\newblock Greenhaven Publishing LLC, 2017.

\bibitem{li2018exposing}
{\sc Li, Y., and Lyu, S.}
\newblock Exposing deepfake videos by detecting face warping artifacts.
\newblock {\em arXiv preprint arXiv:1811.00656 2\/} (2018).

\bibitem{facebookhiring}
{\sc Lyons, T.}
\newblock {Hard Questions: How Is Facebook{'}s Fact-Checking Program Working?
  {$\vert$} Facebook Newsroom}, Sep 2019.
\newblock [Online; accessed 23. Sep. 2019].

\bibitem{wasapBrazilClicks}
{\sc Magenta, M., Gragnani, J., and Souza, F.}
\newblock {WhatsApp weaponised in Brazil election}, Sep 2019.
\newblock [Online; accessed 19. Sep. 2019].

\bibitem{maras2019determining}
{\sc Maras, M.-H., and Alexandrou, A.}
\newblock Determining authenticity of video evidence in the age of artificial
  intelligence and in the wake of deepfake videos.
\newblock {\em The International Journal of Evidence \& Proof 23}, 3 (2019),
  255--262.

\bibitem{marwick2017media}
{\sc Marwick, A., and Lewis, R.}
\newblock Media manipulation and disinformation online.
\newblock {\em New York: Data and Society Research Institute\/} (2017).

\bibitem{McLaughlin2018Dec}
{\sc McLaughlin, T.}
\newblock {How WhatsApp Fuels Fake News and Violence in India}.
\newblock {\em Wired\/} (Dec 2018).

\bibitem{mendoza2010twitter}
{\sc Mendoza, M., Poblete, B., and Castillo, C.}
\newblock Twitter under crisis: Can we trust what we rt?
\newblock In {\em Proceedings of the first workshop on social media
  analytics\/} (2010), ACM, pp.~71--79.

\bibitem{messner2011legitimizing}
{\sc Messner, M., and South, J.}
\newblock Legitimizing wikipedia: How us national newspapers frame and use the
  online encyclopedia in their coverage.
\newblock {\em Journalism Practice 5}, 2 (2011), 145--160.

\bibitem{Kenyansf43:online}
{\sc Miriello, N., Gilbert, D., and Steers, J.}
\newblock Kenyans face a fake news epidemic. - vice.
\newblock
  \url{https://www.vice.com/en_us/article/43bdpm/kenyans-face-a-fake-news-epidemic-they-want-to-know-just-how-much-cambridge-analytica-and-facebook-are-to-blame}.
\newblock (Accessed on 09/18/2019).

\bibitem{jmopatrolling}
{\sc {Morgan, Jonathan}}.
\newblock {Research:Patrolling on Wikipedia/Report - Meta}, Sep 2019.
\newblock [Online; accessed 23. Sep. 2019].

\bibitem{whatisClick}
{\sc Munson, L.}
\newblock {What is a Click Farm?}, Sep 2019.
\newblock [Online; accessed 19. Sep. 2019].

\bibitem{nakashole2014language}
{\sc Nakashole, N., and Mitchell, T.~M.}
\newblock Language-aware truth assessment of fact candidates.
\newblock In {\em Proceedings of the 52nd Annual Meeting of the Association for
  Computational Linguistics (Volume 1: Long Papers)\/} (2014), pp.~1009--1019.

\bibitem{orita20074}
{\sc Orita, A.}
\newblock 4.5 accountable or casual anonymity? a classification of anonymity
  based on linkability.
\newblock In {\em Organizations and Society in Information Systems (OASIS) 2007
  Workshop\/} (2007), p.~33.

\bibitem{pariser2011filter}
{\sc Pariser, E.}
\newblock {\em The filter bubble: What the Internet is hiding from you}.
\newblock Penguin UK, 2011.

\bibitem{penney2016chilling}
{\sc Penney, J.~W.}
\newblock Chilling effects: Online surveillance and wikipedia use.
\newblock {\em Berkeley Tech. LJ 31\/} (2016), 117.

\bibitem{poland2016haters}
{\sc Poland, B.}
\newblock {\em Haters: Harassment, abuse, and violence online}.
\newblock U of Nebraska Press, 2016.

\bibitem{redi2019citation}
{\sc Redi, M., Fetahu, B., Morgan, J., and Taraborelli, D.}
\newblock Citation needed: A taxonomy and algorithmic assessment of wikipedia's
  verifiability.
\newblock In {\em The World Wide Web Conference\/} (2019), ACM, pp.~1567--1578.

\bibitem{resende2018system}
{\sc Resende, G., Messias, J., Silva, M., Almeida, J., Vasconcelos, M., and
  Benevenuto, F.}
\newblock A system for monitoring public political groups in whatsapp.
\newblock In {\em Proceedings of the 24th Brazilian Symposium on Multimedia and
  the Web\/} (2018), ACM, pp.~387--390.

\bibitem{ribeiro2018media}
{\sc Ribeiro, F.~N., Henrique, L., Benevenuto, F., Chakraborty, A.,
  Kulshrestha, J., Babaei, M., and Gummadi, K.~P.}
\newblock Media bias monitor: Quantifying biases of social media news outlets
  at large-scale.
\newblock In {\em Twelfth International AAAI Conference on Web and Social
  Media\/} (2018).

\bibitem{ribeiro2017everything}
{\sc Ribeiro, M.~H., Calais, P.~H., Almeida, V.~A., and Meira~Jr, W.}
\newblock " everything i disagree with is\# fakenews": Correlating political
  polarization and spread of misinformation.
\newblock {\em arXiv preprint arXiv:1706.05924\/} (2017).

\bibitem{ribeiro2019auditing}
{\sc Ribeiro, M.~H., Ottoni, R., West, R., Almeida, V.~A., and Meira, W.}
\newblock Auditing radicalization pathways on youtube.
\newblock {\em arXiv preprint arXiv:1908.08313\/} (2019).

\bibitem{Sampson2015DecCroatia}
{\sc Sampson, T.}
\newblock {How pro-fascist ideologues are rewriting Croatia's history}.
\newblock {\em Daily Dot\/} (Dec 2015).

\bibitem{schafer2017japan}
{\sc Sch{\"a}fer, F., Evert, S., and Heinrich, P.}
\newblock Japan's 2014 general election: Political bots, right-wing internet
  activism, and prime minister shinz{\=o} abe's hidden nationalist agenda.
\newblock {\em Big data 5}, 4 (2017), 294--309.

\bibitem{Schwartz2018Nov}
{\sc Schwartz, O.}
\newblock {You thought fake news was bad? Deep fakes are where truth goes to
  die}.
\newblock {\em the Guardian\/} (Nov 2018).

\bibitem{shang2018investigating}
{\sc Shang, J., Shen, J., Sun, T., Liu, X., Gruenheid, A., Korn, F., Lelkes,
  {\'A}.~D., Yu, C., and Han, J.}
\newblock Investigating rumor news using agreement-aware search.
\newblock In {\em Proceedings of the 27th ACM International Conference on
  Information and Knowledge Management\/} (2018), ACM, pp.~2117--2125.

\bibitem{shi2019wisdom}
{\sc Shi, F., Teplitskiy, M., Duede, E., and Evans, J.~A.}
\newblock The wisdom of polarized crowds.
\newblock {\em Nature human behaviour 3}, 4 (2019), 329.

\bibitem{Shubber2017Oct}
{\sc Shubber, K.}
\newblock {Russia caught editing Wikipedia entry about MH17}.
\newblock {\em WIRED UK\/} (Oct 2017).

\bibitem{spitz2016state}
{\sc Spitz, A., Dixit, V., Richter, L., Gertz, M., and Gei{\ss}, J.}
\newblock State of the union: A data consumer's perspective on wikidata and its
  properties for the classification and resolution of entities.
\newblock In {\em Tenth International AAAI Conference on Web and Social
  Media\/} (2016).

\bibitem{stieglitz2017social}
{\sc Stieglitz, S., Brachten, F., Ross, B., and Jung, A.-K.}
\newblock Do social bots dream of electric sheep? a categorisation of social
  media bot accounts.
\newblock {\em arXiv preprint arXiv:1710.04044\/} (2017).

\bibitem{sumi2011edit}
{\sc Sumi, R., Yasseri, T., et~al.}
\newblock Edit wars in wikipedia.
\newblock In {\em 2011 IEEE Third International Conference on Privacy,
  Security, Risk and Trust and 2011 IEEE Third International Conference on
  Social Computing\/} (2011), IEEE, pp.~724--727.

\bibitem{brazilWhatsApp}
{\sc Tardáguila, C., Benevenuto, F., and Ortellado, P.}
\newblock {Opinion {$\vert$} Fake News Is Poisoning Brazilian Politics.
  WhatsApp Can Stop It.}, Sep 2019.
\newblock [Online; accessed 18. Sep. 2019].

\bibitem{thorne2018fever}
{\sc Thorne, J., Vlachos, A., Christodoulopoulos, C., and Mittal, A.}
\newblock Fever: a large-scale dataset for fact extraction and verification.
\newblock {\em arXiv preprint arXiv:1803.05355\/} (2018).

\bibitem{thorne2018fact}
{\sc Thorne, J., Vlachos, A., Cocarascu, O., Christodoulopoulos, C., and
  Mittal, A.}
\newblock The fact extraction and verification (fever) shared task.
\newblock {\em arXiv preprint arXiv:1811.10971\/} (2018).

\bibitem{tucker2018social}
{\sc Tucker, J.~A., Guess, A., Barber{\'a}, P., Vaccari, C., Siegel, A.,
  Sanovich, S., Stukal, D., and Nyhan, B.}
\newblock Social media, political polarization, and political disinformation: A
  review of the scientific literature.
\newblock {\em Political Polarization, and Political Disinformation: A Review
  of the Scientific Literature (March 19, 2018)\/} (2018).

\bibitem{wardle2017information}
{\sc Wardle, C., and Derakhshan, H.}
\newblock Information disorder: Toward an interdisciplinary framework for
  research and policy making.
\newblock {\em Council of Europe Report 27\/} (2017).

\bibitem{Wikipedi57:online}
{\sc Wikipedia}.
\newblock Wikipedia:edit warring.
\newblock \url{https://en.wikipedia.org/wiki/Wikipedia:Edit_warring}.
\newblock (Accessed on 09/18/2019).

\bibitem{Wikipedi81:neutral}
{\sc Wikipedia}.
\newblock Wikipedia:neutral point of view.
\newblock \url{https://en.wikipedia.org/wiki/Wikipedia:Neutral_point_of_view}.
\newblock (Accessed on 09/18/2019).

\bibitem{Wikipedi36:online}
{\sc Wikipedia}.
\newblock Wikipedia:no original research.
\newblock \url{https://en.wikipedia.org/wiki/Wikipedia:No_original_research}.
\newblock (Accessed on 09/18/2019).

\bibitem{Wikipedi87:veri}
{\sc Wikipedia}.
\newblock Wikipedia:verifiability.
\newblock \url{https://en.wikipedia.org/wiki/Wikipedia:Verifiability}.
\newblock (Accessed on 09/18/2019).

\bibitem{israelWikipedia}
{\sc Wikipedia}.
\newblock {Committee for Accuracy in Middle East Reporting in America}, Sep
  2019.
\newblock [Online; accessed 20. Sep. 2019].

\bibitem{orangemoody}
{\sc Wikipedia}.
\newblock {Orangemoody editing of Wikipedia - Wikipedia}, Sep 2019.
\newblock [Online; accessed 20. Sep. 2019].

\bibitem{bulgaria}
{\sc Wikipedia}.
\newblock {Proposals for closing projects/Deletion of Bulgarian Wikinews -
  Meta}, Sep 2019.
\newblock [Online; accessed 21. Sep. 2019].

\bibitem{azwiki}
{\sc Wikipedia}.
\newblock {Requests for comment/Do something about azwiki - Meta}, Sep 2019.
\newblock [Online; accessed 20. Sep. 2019].

\bibitem{scientology}
{\sc Wikipedia}.
\newblock {Scientology}, Sep 2019.
\newblock [Online; accessed 20. Sep. 2019].

\bibitem{sockpuppets}
{\sc Wikipedia}.
\newblock {Sockpuppet (Internet)}, Sep 2019.
\newblock [Online; accessed 19. Sep. 2019].

\bibitem{sybilAtaccks}
{\sc {Wikipedia}}.
\newblock {Sybil attack}, Sep 2019.
\newblock [Online; accessed 19. Sep. 2019].

\bibitem{citingWikipeda}
{\sc {Wikipedia}}.
\newblock {Wikipedia:Academic use }, Sep 2019.
\newblock [Online; accessed 21. Sep. 2019].

\bibitem{circularList}
{\sc Wikipedia}.
\newblock {Wikipedia:List of citogenesis incidents}, Sep 2019.
\newblock [Online; accessed 21. Sep. 2019].

\bibitem{sockPupetry}
{\sc Wikipedia}.
\newblock {Wikipedia:Sock puppetry }, Oct 2019.
\newblock [Online; accessed 14. Oct. 2019].

\bibitem{lessons1942}
{\sc Zeitz, J.}
\newblock {Lessons From the Fake News Pandemic of 1942}, Sep 2019.
\newblock [Online; accessed 18. Sep. 2019].

\bibitem{zhang2013online}
{\sc Zhang, J., Carpenter, D., and Ko, M.}
\newblock Online astroturfing: A theoretical perspective.

\bibitem{zhou2018fake}
{\sc Zhou, X., and Zafarani, R.}
\newblock Fake news: A survey of research, detection methods, and
  opportunities.
\newblock {\em arXiv preprint arXiv:1812.00315\/} (2018).

\end{thebibliography}


\end{document}